\documentclass[12pt]{iopart}
\usepackage{graphicx}
\usepackage{bm}

\begin{document}
\title{Additional resonant contribution to the potential model for the $^{12}$C($\alpha$,$\gamma$)$^{16}$O reaction}
\author{M.~Katsuma}
\address{Advanced Mathematical Institute, Osaka City University, Osaka 558-8585, Japan}
\ead{mkatsuma@sci.osaka-cu.ac.jp}

\begin{abstract}
  The additional resonant contribution to the potential model is examined in $\alpha$+$^{12}$C elastic scattering and the low-energy $^{12}$C($\alpha$,$\gamma$)$^{16}$O reaction.
  The excitation function of elastic scattering below $E_{c.m.}= 5$ MeV seems to be reproduced by the potential model satisfactorily, and it is not profoundly disturbed by the additional resonances.
  The weak coupling is good enough to describe the $^{16}$O structure in the vicinity of the $\alpha$-particle threshold, especially below $E_{c.m.}= 8$ MeV, corresponding to the excitation energy $E_x \approx 15$ MeV.
  The additional resonances give the complement of the astrophysical $S$-factors from the simple potential model.
  The $S$-factor of $^{12}$C($\alpha$,$\gamma$)$^{16}$O at $E_{c.m.}=300$ keV is dominated by the $E$2 transition, which is enhanced by the subthreshold 2$^+_1$ state at $E_x= 6.92$ MeV.
  The contribution from the subthreshold 1$^-_1$ state at $E_x= 7.12$ MeV is predicted to be small.
  The additional resonances do not give the large contribution to the thermonuclear reaction rates of $^{12}$C($\alpha$,$\gamma$)$^{16}$O at helium burning temperatures.
\end{abstract}

\submitto{\JPG}
\maketitle

\section{Introduction}
  The $^{12}$C($\alpha$,$\gamma$)$^{16}$O reaction following the triple $\alpha$ reaction in stars plays the very important role in the production of heavier nuclei than carbon \cite{Rol88}.
  To scrutinize the origin of elements, the low-energy $^{12}$C($\alpha$,$\gamma$)$^{16}$O cross sections have been investigated at helium burning temperatures, corresponding to the center-of-mass energy $E_{c.m.}\approx 300$ keV.
  However, the cross sections are very small, owing to the Coulomb barrier, so the direct measurement is not feasible at the present laboratories.

  To cope with the difficulty, the theoretical model calculation has been performed with the simple potential model \cite{Kat12,Kat08}.
  In this model, $\alpha$+$^{12}$C elastic scattering has been scrutinized to illustrate the feature of the $\alpha$+$^{12}$C continuum state.
  Below $E_{c.m.}=5$ MeV, the elastic cross sections have been found to be described very well by the simple $\alpha$+$^{12}$C configuration \cite{Kat08,Kat10}.
  The resulting potential between $\alpha$-particle and $^{12}$C nuclei is concordant with the optical model potential reproducing elastic scattering at laboratory energies $E_\alpha \approx 100$ MeV where the ambiguity of the potential is eliminated \cite{Ing94,Nol87,Mic95,Mic83,Bra97,Sat83}.
  The $\alpha$+$^{12}$C rotational bands are well reproduced and the 8$^+$ and 9$^-$ states at the excitation energy $E_x\approx$ 30 MeV are predicted to be the known rotational band member \cite{Kat14,Kat13}.
  From the characteristic feature of the reaction mechanism and the $^{16}$O structure, the low-energy $^{12}$C($\alpha$,$\gamma$)$^{16}$O cross sections have been calculated and they have been converted into the astrophysical reaction rates \cite{Kat12,Kat08}.
  At $E_{c.m.}= 300$ keV, the radiative capture cross section is dominated by the $E$2 transition to the ground state.
  The cascade transitions are important above $E_{c.m.}= 1$ MeV, corresponding to $T_9\approx 1$.
  $T_9$ is the temperature in the unit of $T_9=10^9$ K.
  The microscopic models have also attempted to describe the $^{12}$C($\alpha$,$\gamma$)$^{16}$O reaction and the structure of $^{16}$O (e.g. \cite{Duf08}), theoretically.

  Recently, the devoted efforts of the progress in the experimental work have been made, as well as the theoretical predictions.
  At low energies, the $\gamma$-ray angular distribution and its ambiguities have been discussed \cite{Gai13,Pla12,Mak09,Ass06,Kun01}.
  To cultivate the knowledge of $^{12}$C($\alpha$,$\gamma$)$^{16}$O, the direct measurement of cross sections, the cascade transition through the excited states of $^{16}$O and the total capture reaction cross sections have been investigated experimentally (e.g. \cite{Sch12,Sch11,Sch05,Mat06,Kun02,Rot99,Red87,Ket82}).
  However, these measurable energies correspond to the relatively high temperatures, even though they use the current technologies.
  So, the extrapolated values are made by e.g. the $R$-matrix method \cite{Lan58}.
  To pave the way for the analyses, the phase shifts of $\alpha$+$^{12}$C elastic scattering have also been measured precisely \cite{Tis09,Tis02,Pla87}.
  The indirect measurements (e.g. \cite{Bru99,Bel07}) and the $\beta$-delayed $\alpha$ decay of $^{16}$N (e.g. \cite{Tan10,Buc06,Buc96,Azu94,Buc93}) have been demonstrated to evaluate the $\alpha$-particle width of the subthreshold 1$^-_1$ state ($E_x= 7.12$ MeV), alternatively.

  If the radiative capture cross sections are expressed by the Breit-Wigner form, the high energy side of the tail of the subthreshold state could contribute the enhancement of the reaction rates \cite{Rol88,Nacre}.
  The low-energy $E$1 cross sections are believed to be enhanced by the 1$^-_1$ state.

  Above $E_{c.m.}\simeq 5$ MeV, the $^{15}$N(p,$\alpha$)$^{12}$C and the $^{15}$N(p,$\gamma$)$^{16}$O reactions are available (e.g.~\cite{Cog07,Cog09,Bar08,Mar10,Imb12, Imb12e,Rol74,Nacre}).
  The branching ratio of these reactions determines the escape from the main CNO cycle to the CNO-II cycle, and it controls the energy production of the proton burning in a star.
  The low-lying two 1$^-$ resonant states ($E_x=12.44$ MeV and 13.09 MeV) in the p+$^{15}$N channel appear to be coupled to the $\alpha$+$^{12}$C continuum state.
  Even at higher energies, the study of the $\alpha$+$^{12}$C system is important in nuclear astrophysics.

  In our previous studies \cite{Kat12,Kat08}, we have investigated the $^{12}$C($\alpha$,$\gamma$)$^{16}$O reaction at sub-barrier energies (below $E_{c.m.}=3$ MeV), and we have provided the derived reaction rates below $T_9=3$.
  Above the barrier, we see the narrow resonances in the excitation function of elastic scattering (e.g. \cite{Til93,Mar72}) and $^{12}$C($\alpha$,$\gamma$)$^{16}$O (e.g. \cite{Sch12,Sch05}).
  The contribution from the resonances has not been explicitly discussed yet. 
  At $T_9=3$, the so-called Gamov peak energy $E_0$ and width $\Delta E_0$ \cite{Rol88,Nacre} are $E_0=1.92$ MeV and $\Delta E_0=1.63$ MeV, respectively.
  If taking account of the numerical integration up to $E_0+3\Delta E_0=6.8$ MeV \cite{Nacre}, we might want to discuss the contribution from the cross sections above the barrier.

  In the present article, we investigate the additional resonant contribution to the potential model for $\alpha$+$^{12}$C elastic scattering and the $^{12}$C($\alpha$,$\gamma$)$^{16}$O reaction.
  We show whether the potential model reproduces the excitation function of $\alpha$+$^{12}$C elastic scattering and the astrophysical $S$-factors, including the known resonances.
  We estimate the difference in the derived reaction rates below $T_9=3$.
  The contribution from the subthreshold states is also discussed.
  The main purpose of the present study is to examine the additional resonant contribution to the calculated cross sections and reaction rates.

  In the following section, we explain the potential model with the additional resonances.
  In Section \ref{sec3}, the contribution from the additional resonances is discussed by showing the difference in the excitation function of $\alpha$+$^{12}$C elastic scattering and the $^{12}$C($\alpha$,$\gamma_0$)$^{16}$O $S$-factors.
  The derived reaction rates are also compared with the previous ones \cite{Kat12}.
  The summary is given in Section \ref{sec4}.

\section{Potential model with the additional resonances}
\label{sec2}

  In this section, we describe the potential model \cite{Kat12,Kat08,Des,Ros67}.
  The basic idea of the potential model is based on the mean field approximation of quantum many-body systems \cite{Sat83}.
  The potential model describes relative motion between interacting two nuclei, and it makes the potential resonance and the smooth variation on energy in the excitation function.
  The narrow resonances are appended to the result obtained from the simple potential model.
  We refer the simple potential model calculation without the additional resonances as the direct-capture component or potential scattering in the present article.
  The additional resonant term is called the resonant component or dynamical process, because it originates from the couplings to other reaction channels.

  The astrophysical $S$-factors for the radiative capture reaction are conventionally used instead of the cross sections to compensate for the rapid drop of the cross sections below the barrier \cite{Rol88,Nacre,Fow67}, and they are defined by
  \begin{eqnarray}
     S_{E\lambda}(E_{c.m.}) &=& E_{c.m.} \exp(2\pi\eta)\, \sigma_{E\lambda}(E_{c.m.}),
     \label{eq:sfact}
  \end{eqnarray}
  where $\eta$ is the Sommerfeld parameter, $\eta=Z_1Z_2e^2/(\hbar v)$; $v=\sqrt{2E_{c.m.}/\mu}$; $Z_1$ and $Z_2$ are the charge of $\alpha$-particle and $^{12}$C; $\mu$ is the reduced mass.
  $\lambda$ is multipole of the transition.
  $\sigma_{E\lambda}(E_{c.m.})$ is the radiative capture cross section, defined by
  \begin{eqnarray}
    \sigma_{E\lambda}(E_{c.m.}) 
    &=&
    \mathop{\sum}_f
    \frac{2J_f+1}{(2I_1+1)(2I_2+1)} \frac{\pi}{k_i^2} 
    \mathop{\sum}_{l_i} 
    \big|\, T^{E\lambda}_{l_fl_i}(k_f,k_i) \, \big|^2.
    \label{eq:xsec}
  \end{eqnarray}
  $J_f$ is the spin of the final bound state; $I_1$ and $I_2$ are the spin of interacting two nuclei.
  $k$ and $l$ are the wave number and the angular momentum of relative motion between $\alpha$ and $^{12}$C.
  The subscript denotes the value for the initial channel ($i$) and final state ($f$).

  The transition amplitudes $T^{E\lambda}_{l_fl_i}(k_f,k_i)$ is defined by
  \begin{eqnarray}
    T^{E\lambda}_{l_fl_i} (k_f,k_i)
    &=&
    \bar{T}^{E\lambda}_{l_fl_i} (k_f,k_i)
    -i\mathop{\sum}_n \tilde{e}_{E\lambda} \frac{\sqrt{\it \Gamma_{\gamma n} \Gamma_{\alpha n}}}{E_{c.m.}-E_n+i{\it \Gamma}_n/2},
    \label{eq:tmat}
  \end{eqnarray}
  where $E_n$, ${\it \Gamma}_{\alpha n}$, ${\it \Gamma}_{\gamma n}$ and ${\it \Gamma_n}$ are the resonance energy, the $\alpha$-particle width, the $\gamma$-width and total width of the resonance $n$, respectively.
  The first term represent the direct-capture component.
  For $^{12}$C($\alpha$,$\gamma_0$)$^{16}$O, the direct-capture component is given by 
  \begin{eqnarray}
    \bar{T}^{E\lambda}_{0 l_i} (k_f,k_i)
    &=&
    \tilde{e}_{E\lambda} C_\lambda \left[\frac{8\pi(\lambda+1)k_\gamma^{2\lambda+1}}{\hbar v \lambda[(2\lambda+1)!!]^2} \right]^{1/2} 
    \int \chi_{f,0}(k_fr) r^\lambda \chi_{i,l_i}(k_ir) dr,
    \label{eq:tmat-direct}
  \end{eqnarray}
  where $C_\lambda$ is the geometric factor.
  $\tilde{e}_{E\lambda}$ is the effective charge for $E\lambda$ transition, which is determined from the direct-capture component.
  $k_\gamma$ is the wave number of the emitted $\gamma$-rays, $k_\gamma=(E_{c.m.}-E_f)/(\hbar c)$.
  $E_f$ is the binding energy of the final state.
  $\chi_{i,f}$ is the wavefunction of relative motion obtained from solving the Schr\"odinger equation with the local potential.
  We adopt the parity-dependent nuclear potential,
  \begin{eqnarray}
      V(r) &=& \left\{ \begin{array}{cc}
            V_+ f_+(r) & {\rm (even)} \\
            V_- f_-(r) & {\rm (odd)} \\
           \end{array}, \right.
      \label{eq:v}
      \\
      f_\xi(r) &=& \frac{1}{1+\exp\,[\,(r-R_\xi)/a_\xi\,]}.
      \label{eq:ws}
  \end{eqnarray}
  $V_\xi$, $R_\xi$ and $a_\xi$ are the potential parameters for the even ($+$) and odd ($-$) parities.
  We use the same parameters as those in the previous studies \cite{Kat12,Kat08,Kat10}: $V_+=-199.7$ MeV, $R_+= 2.18$ fm, $a_+= 0.743$ fm, $V_-=-168.1$ MeV, $R_-= 2.76$ fm, and $a_-= 0.567$ fm for $\alpha$+$^{12}$C continuum state.
  The $V_+$ is adjusted to reproduce the $\alpha$-particle separation energy for the ground state.
  The Coulomb potential is calculated from the uniform charge sphere with a radius $R_C=$ 3.5 fm.
  The $E$1 and $E$2 effective charges are $\tilde{e}_{E1}=9.96\times10^{-3} e$ and $\tilde{e}_{E2}= 1.69 e$ \cite{Kat08}.
  $\tilde{e}_{E1}=0.68 e$ \cite{Kat08} is used for the resonance at $E_n= 5.928$ MeV.

  The nuclear reactions proceed in the sequence of elastic scattering, direct reactions and more complicated nuclear reactions \cite{Sat83}.
  The direct-capture component describes the fundamental process of the radiative capture reactions.
  The more complicated reaction process occurs after the direct-capture process with penetrating the barrier.
  In equation~(\ref{eq:tmat}), the Breit-Wigner form is attached to include the dynamical process.
  The included resonances are listed in Table~\ref{tb:bw}.
  The 0$^+$($E_{c.m.}=-1.113$ MeV), 1$^-$($E_{c.m.}= 2.42$ MeV), 2$^+$($E_{c.m.}=-0.245$ MeV), 3$^-$($E_{c.m.}= 4.44$ MeV), 4$^+$($E_{c.m.}= 3.19$ MeV) and 5$^-$($E_{c.m.}= 7.50$ MeV) states are not included as the additional resonances.
  These are generated by the potential resonance.

  In the present work, we do not perform the $\chi^2$ optimization to the experimental data.
  The purpose of the present article is to illustrate the contribution from the additional resonances, in comparison with the direct-capture component of our previous studies \cite{Kat12,Kat08}.

  For elastic scattering, we utilize the dispersion formula of the $S$-matrix.
  \begin{eqnarray}
    {\cal S}_L &=& \bar{{\cal S}}_L -i \mathop{\sum}_n \frac{{\it \Gamma}_{\alpha n} \e^{i\theta_{\alpha n}}}{E_{c.m.}-E_n+i{\it \Gamma}_n/2},
    \label{eq:smat}
  \end{eqnarray}
  where $\bar{\cal S}_L$ is the $S$-matrix from potential scattering.
  The second term is the dynamical component.
  $\theta_{\alpha n}= 2\bar{\delta}_{l_i}$ is used. $\bar{\delta}_{l_i}$ is the phase shifts of potential scattering.
  The excitation function of elastic scattering is calculated from equation~(\ref{eq:smat}).

  From the energy dependence of wavefunctions at low energies, the $\alpha$-particle width is assumed to be given by attaching a factor to the energy-independent experimental value \cite{Rol88,Nacre},
  \begin{eqnarray}
    {\it \Gamma}_{\alpha n}(E_{c.m.}) 
    &=& \frac{P(E_{c.m.})}{P(|E_n|)}\,{\it \Gamma}_{\alpha n}.
    \label{eq:Gamma-a}
  \end{eqnarray}
  $P(E_{c.m.})$ is defined by the Gamov factor,
  \begin{eqnarray}
    P(E_{c.m.}) &=& \exp\big[-2\pi\eta(E_{c.m.})\big].
    \label{eq:pene-low}
  \end{eqnarray}
  $\eta(|E_n|)=-\eta(|E_n|)$ is assumed to be used for the negative energy.
  The energy dependence of $\gamma$-width comes from the wavelength of the emitted $\gamma$-rays,
  \begin{eqnarray}
    {\it \Gamma}_{\gamma n}(E_{c.m.}) &=& \left(\frac{E_{c.m.}-E_f}{E_n-E_f} \right)^{2\lambda+1} {\it \Gamma}_{\gamma n}.
    \label{eq:Gamma-g}
  \end{eqnarray}

  The Maxwellian-averaged reaction rates $N_A \langle \sigma v \rangle$ \cite{Rol88,Nacre,Fow67} are converted from the calculated $S$-factors.
  \begin{eqnarray}
    N_A \langle \sigma v \rangle 
    &=&
    N_A \frac{(8/\pi)^{1/2}}{\mu^{1/2}(k_BT)^{3/2}} \int \sigma_{E\lambda}(E_{\rm c.m.}) E_{\rm c.m.} 
    \exp\left(-\frac{E_{\rm c.m.}}{k_BT}\right) dE_{\rm c.m.}, 
    \label{eq:rate}
  \end{eqnarray}
  where $N_A$, $k_B$, and $T$ are the Avogadro number, Boltzmann constant and temperature, respectively.
  The $S$-factors consist of the direct-capture and dynamical components.
  So, the additional resonances increase the reaction rates from the previous ones \cite{Kat12}.

  In addition, we estimate the reaction rates from the resonances by using the formula \cite{Rol88,Nacre,Fow67},
  \begin{eqnarray}
    N_A \langle \sigma v \rangle_R
    &=&
    N_A \left(\frac{2\pi}{\mu k_B}\right)^{3/2} \hbar^2 \mathop{\sum}_n (\omega \gamma)_n T^{-3/2} \exp(-E_n/k_B T) +N_A \left(\frac{2}{\mu}\right)^{1/2} \nonumber\\
    &\cdot&
    \frac{\Delta E_0}{(k_B T)^{3/2}} \mathop{\sum}_n S_n(E_0) 
    \exp\left[-3\Big(\frac{\pi^2\mu}{2k_BT}\Big)^{1/3} \Big(\frac{e^2Z_1Z_2}{\hbar}\Big)^{2/3} \right],
    \label{eq:rate-res}
  \end{eqnarray}
  where $(\omega \gamma)_n$ is given by
  \begin{eqnarray}
    (\omega \gamma)_n &=& \frac{\it \Gamma_{\gamma_0 n}\Gamma_{\alpha n}}{\it \Gamma_n}.
    \label{eq:wg}
  \end{eqnarray}
  $S_n(E_0)$ is the $S$-factor of the resonance $n$.
  $E_0$ and $\Delta E_0$ are given by $E_0 \approx 0.922 T_9^{2/3}$ and $\Delta E_0\approx 0.651 T_9^{5/6}$, respectively.
  The formula of equation~(\ref{eq:rate-res}) is utilized in general to estimate roughly the reaction rates although it does not include the interference between the resonances.
  When it is compared with the numerical integration of equation~(\ref{eq:rate}), the direct-capture component is added to equation~(\ref{eq:rate-res}).

\section{Results}
\label{sec3}

  In this section, we examine the additional resonant contribution to the potential model for $\alpha$+$^{12}$C elastic scattering and $^{12}$C($\alpha$,$\gamma$)$^{16}$O.
  We compare the calculated results of the excitation function and the phase shifts with those \cite{Kat10} from potential scattering below $E_{c.m.}= 5$ MeV.
  After illustrating the applicability of the model, we discuss the effect to the astrophysical $S$-factor of the $^{12}$C($\alpha$,$\gamma$)$^{16}$O reaction.
  Using the $E$2 transition, we recall the tail contribution of the subthreshold state in the $S$-factors.
  We explain the contribution from the additional subthreshold 1$^-_1$ state, which is not included explicitly in the potential model with the simple $\alpha$+$^{12}$C configuration.
  We finally estimate the additional contribution to the derived reaction rates below $T_9=3$.

\subsection{$\alpha$+$^{12}$C elastic scattering}

\begin{figure}
  \begin{center}
    \includegraphics[width=0.48\linewidth]{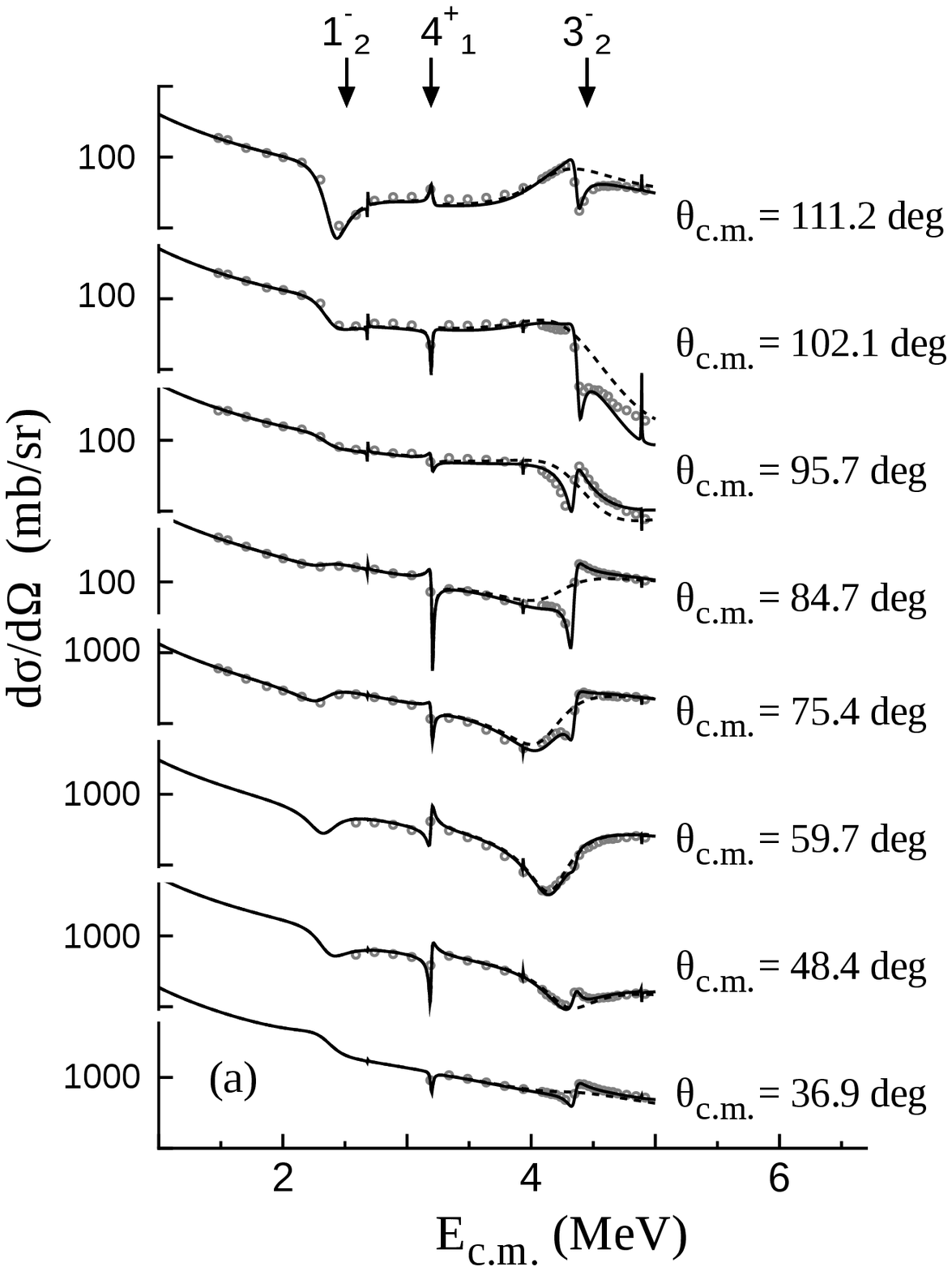}
    \includegraphics[width=0.48\linewidth]{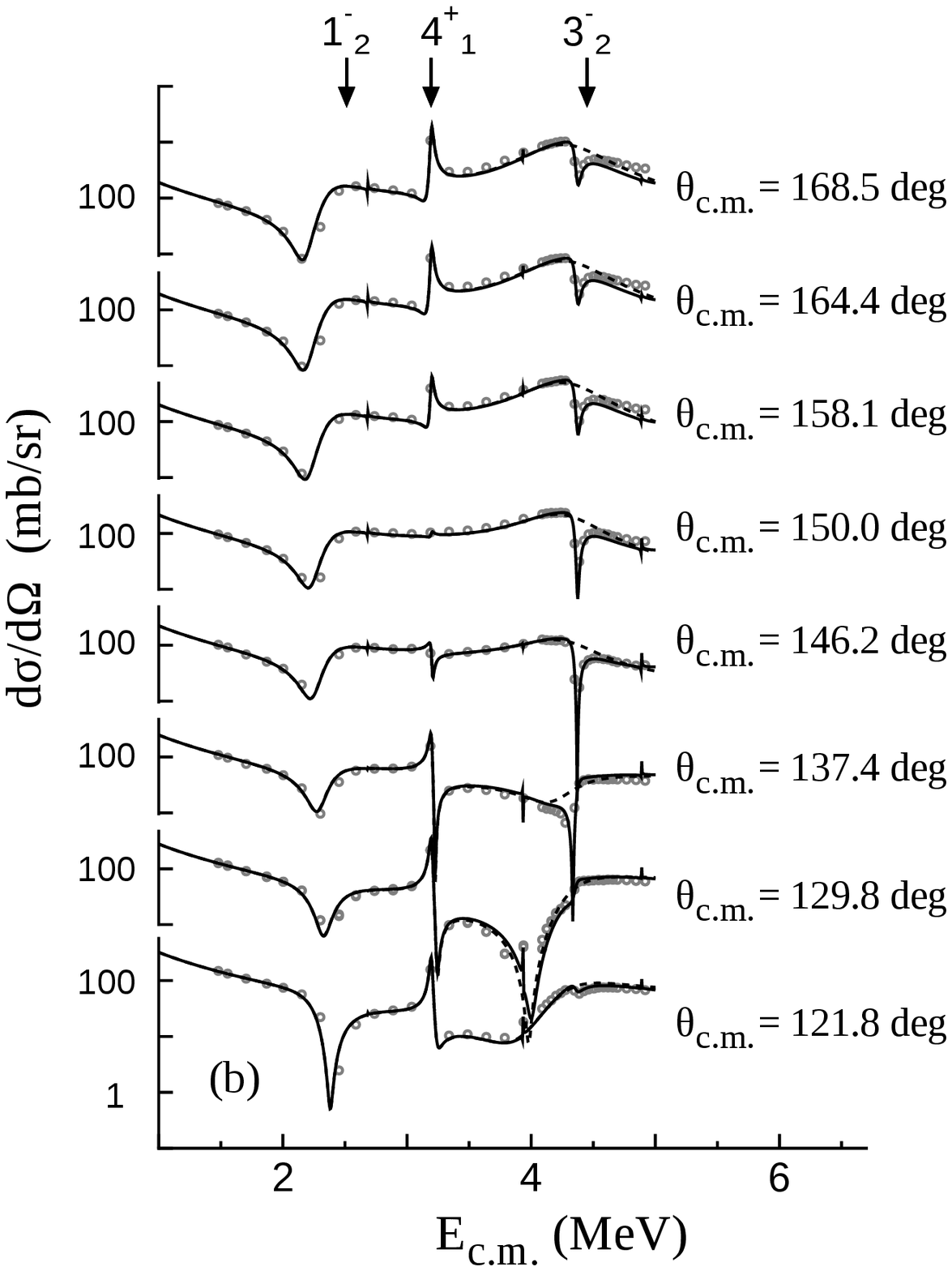}
  \end{center}
  \caption{\label{fig:ex1}
    The excitation function of $\alpha$+$^{12}$C elastic scattering below $E_{c.m.}=5$ MeV.
    The solid curves are the results obtained from the potential model with the additional resonances.
    The dotted curves represent the results from potential scattering.
    The energy position of the $\alpha$+$^{12}$C rotational band \cite{Kat14,Kat13} is shown by the arrows.
    The scattering angles are displayed besides the respective curves.
    The experimental data are taken from \cite{Pla87,EXFOR}.
  }
\end{figure}

  Figure \ref{fig:ex1} shows the excitation function of $\alpha$+$^{12}$C elastic scattering below $E_{c.m.}= 5$ MeV for (a) $\theta_{c.m.}=36.9^\circ$ -- 111.2$^\circ$ and (b) $\theta_{c.m.}=121.8^\circ$ -- 168.5$^\circ$.
  The solid curves are the calculated results with the additional resonances.
  The dotted curves are obtained from potential scattering \cite{Kat10}.
  At $E_{c.m.}\approx 4.4$ MeV, we see the difference between two curves due to the 2$^+_3$ resonance at $E_{c.m.}=4.357$ MeV ($E_x=11.52$ MeV).
  We also see the small peak at $E_{c.m.}\approx 2.683$ MeV, which originates from the 2$^+_2$ narrow resonance.
  It is, however, found that the additional resonances give no more than the slight deviation from potential scattering.
  The arrows indicate the energy position of the states belonging to the rotational bands \cite{Kat14,Kat13}.
  From the comparison between two curves, the weak coupling is confirmed to be good enough to describe the excitation function below $E_{c.m.} = 5$ MeV.

\begin{figure}
  \begin{center}
    \includegraphics[width=0.5\linewidth]{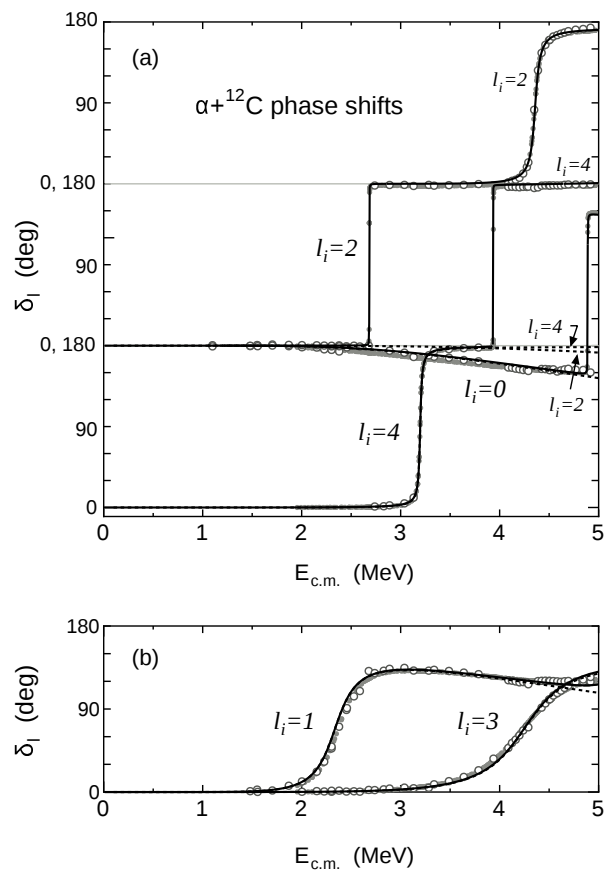}
  \end{center}
  \caption{\label{fig:ps1}
    The phase shifts of (a) $l_i=0$, 2 and 4 (b) $l_i=1$ and 3 for $\alpha$+$^{12}$C elastic scattering.
    The solid curves are the calculated phase shifts with the additional resonances.
    The dotted curves represent the results from potential scattering.
    The experimental data are taken from \cite{Pla87,Tis09}.
  }
\end{figure}

  The corresponding phase shifts for (a) $l_i=0, 2, 4$ and (b) $l_i=1, 3$ are shown in figure~\ref{fig:ps1}.
  The solid and dotted curves are the results with/without the additional resonances.
  The $l_i=2$ phase shift has the step of 180$^\circ$ because of the 2$^+_2$ ($E_{c.m.}= 2.683$ MeV) and 2$^+_3$ ($E_{c.m.}= 4.357$ MeV) resonances.
  However, the difference of the $l_i=2$ phase shift is localized around the resonance energies because $\delta_{l_i}=0$ is equal to $\delta_{l_i}=180$.
  The difference in the phase shifts reflects to the deviation between two curves of the excitation function shown in figure~\ref{fig:ex1}.
  We figure out that the additional resonances do not contribute to elastic scattering below $E_{c.m.}= 5$ MeV, substantially.

\begin{figure}
  \begin{center}
    \includegraphics[width=0.49\linewidth]{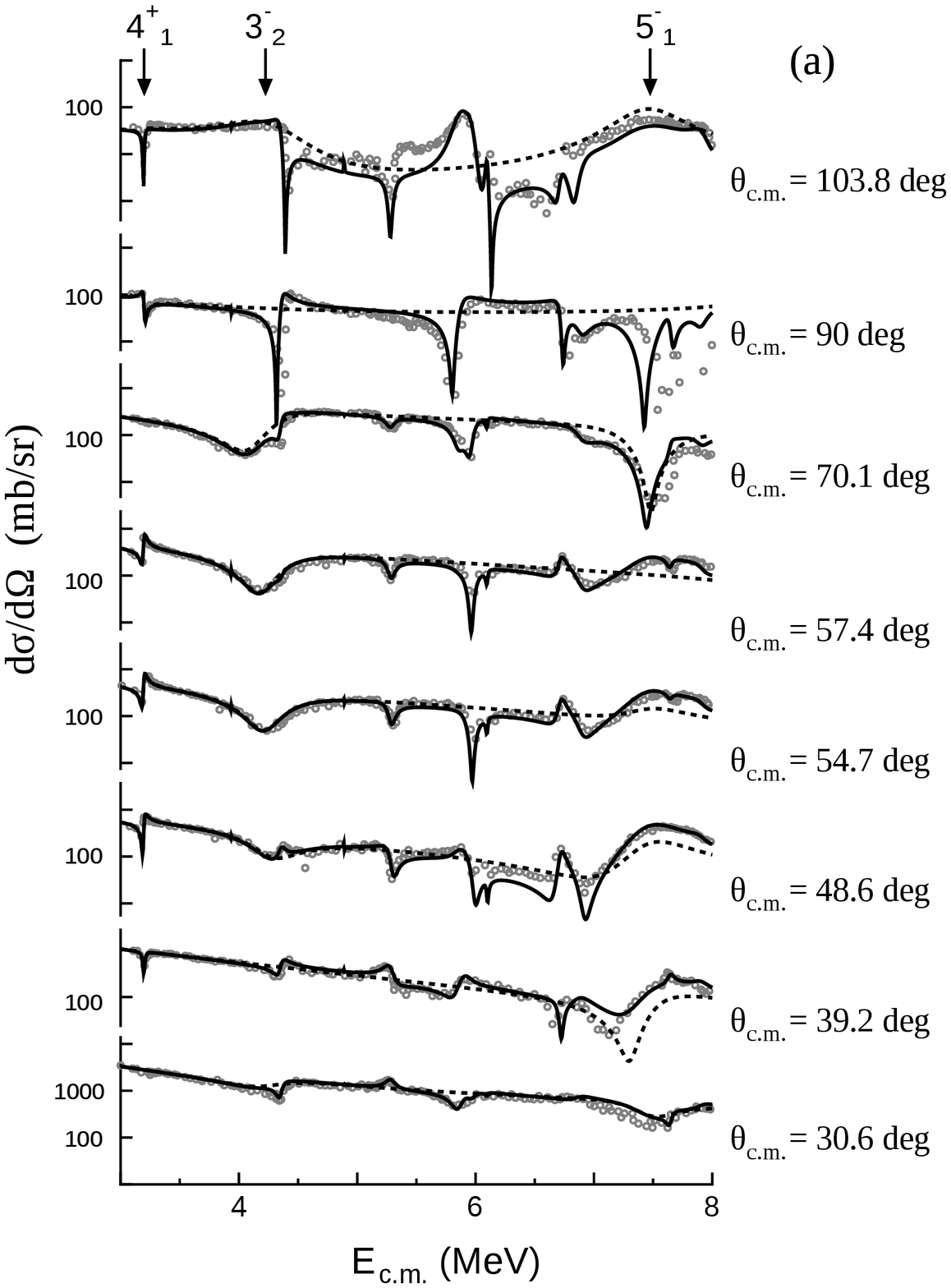}
    \includegraphics[width=0.49\linewidth]{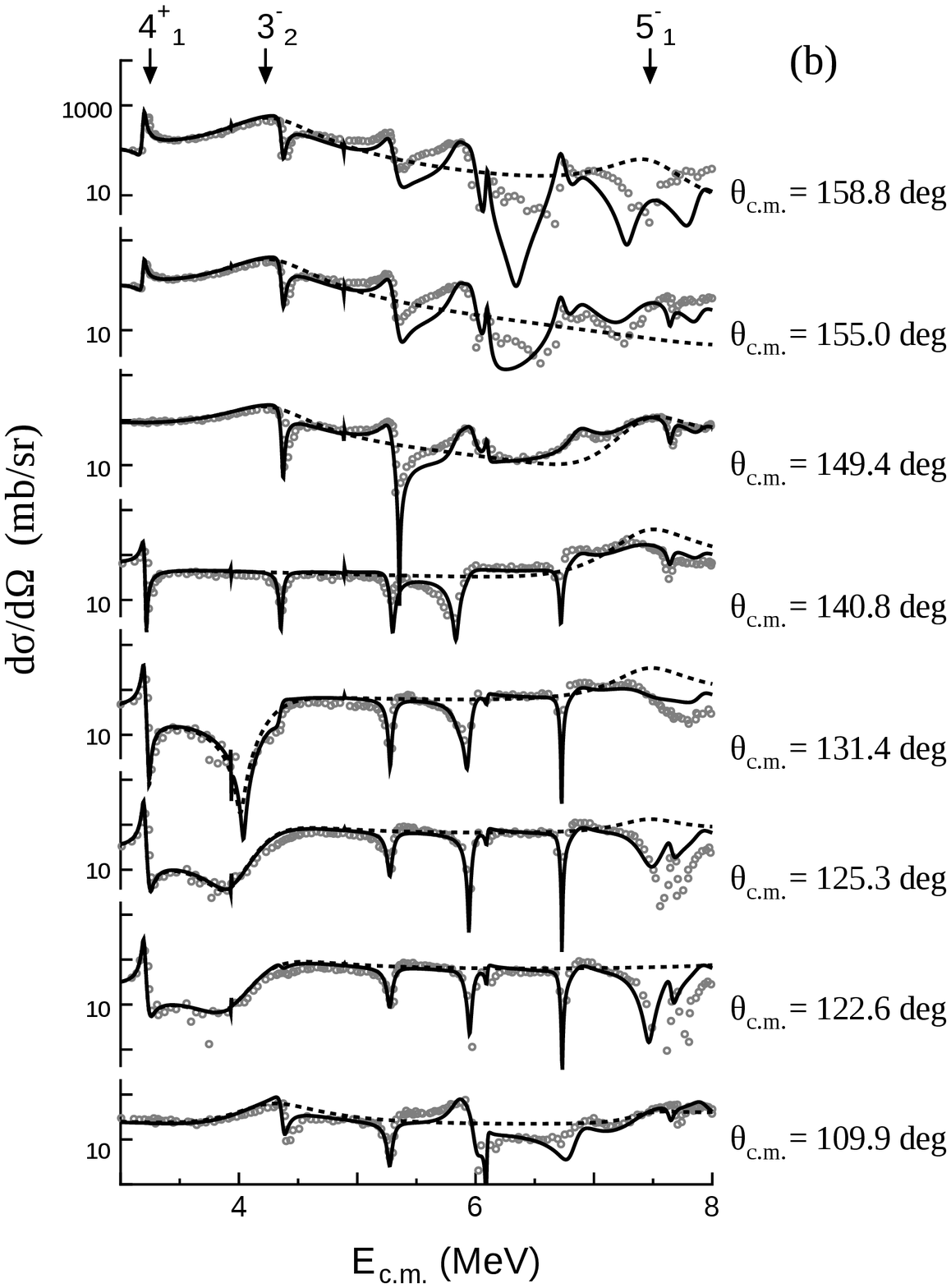}
  \end{center}
  \caption{\label{fig:ex2}
    The excitation function of $\alpha$+$^{12}$C elastic scattering in the energy region of $E_{c.m.}=3$ -- 8 MeV.
    The solid curves are the results obtained from the potential model with the additional resonances.
    The dotted curves represent the results from potential scattering.
    The energy position of the $\alpha$+$^{12}$C rotational band \cite{Kat14,Kat13} is shown by the arrows.
    The scattering angles are displayed besides the respective curves.
    The experimental data are taken from \cite{Mar72,EXFOR}.
  }
\end{figure}

  To verify that the present model is suitable above the barrier, we illustrate the excitation function of $\alpha$+$^{12}$C elastic scattering up to $E_{c.m.}= 8$ MeV, as well.
  The solid and dotted curves in figure~\ref{fig:ex2} are the results obtained from the potential model with/without the additional resonances.
  The characteristic feature of the experimental data appears to be reproduced by the solid curves.
  The potential scattering shown by the dotted curves seems to give the smooth trend of the excitation function and the potential resonances.
  The remaining rapid variations come from the additional resonances (Table~\ref{tb:bw}).
  The strength of the potential for $l_i=5$ is slightly adjusted as $V_{l=5}=-170.3$ MeV so as to reproduce the experimental $l_i=5$ resonance.
  From the comparison between two curves, we find that the weak coupling is still appropriate for describing the outline of the $\alpha$+$^{12}$C excitation function up to $E_{c.m.}=8$ MeV.
   We here confirm that the excited states of $^{16}$O nuclei below $E_x\approx 15$ MeV are described by the present model.

\subsection{Astrophysical $S$-factors}

\begin{figure}
  \begin{center}
    \includegraphics[width=0.7\linewidth]{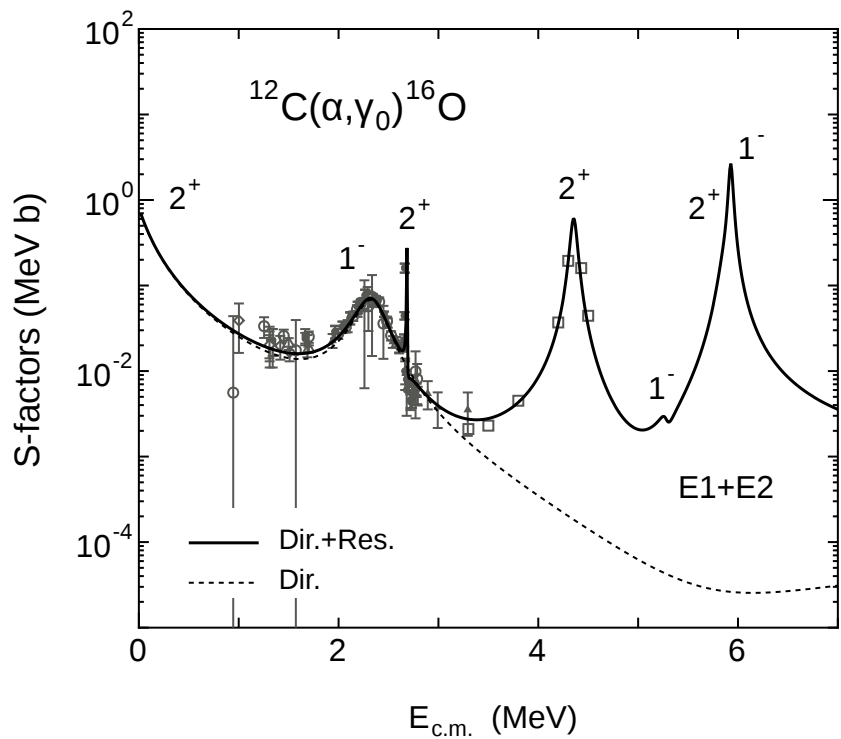}
  \end{center}
  \caption{\label{fig:sfact}
    The astrophysical $S$-factors of the $^{12}$C($\alpha$,$\gamma_0$)$^{16}$O reaction.
    The solid curve is the result obtained from the potential model with the additional resonances.
    The dotted curve is the direct-capture component.
    The $S$-factors are the sum of the $E$1 and $E$2 transition to the ground state of $^{16}$O.
    The experimental data are taken from \cite{Ass06,Kun01,Mak09,Pla12,Rot99,Sch12,EXFOR}.
  }
\end{figure}

  The calculated $S$-factors are compared in figure~\ref{fig:sfact}.
  The solid curve is the result obtained from the potential model with the additional resonances, and it is the sum of the $S$-factors of the $E$1 and $E$2 transitions to the ground state of $^{16}$O.
  The dotted curves are the direct-capture component.
  We see the large difference because of the resonances at $E_{c.m.}=4.357$ MeV, 5.865 MeV and 5.928 MeV.
  The narrow resonance makes the sharp peak at $E_{c.m.}=2.683$ MeV.
  We, however, find that the $S$-factors are not disturbed by the additional resonances below $E_{c.m.}= 3$ MeV.
  The small deviation makes us expect that the additional resonances do not give the profound change of the reaction rates below $T_9\approx 3$.
  The $S$-factors at low energies are dominated by the $E$2 transition in the potential model, as discussed in \cite{Kat12,Kat08}.

\begin{figure}
  \begin{center}
    \includegraphics[width=0.6\linewidth]{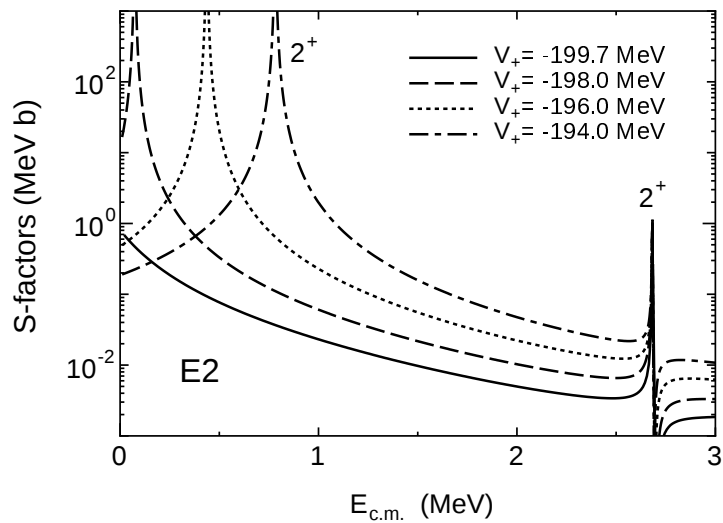}
  \end{center}
  \caption{\label{fig:E2}
    The schematic calculation of the $E$2 $S$-factors.
    The $V_+$ is varied to reveal the reason of the enhanced $E$2 $S$-factors.
    The energy position of the 2$^+_1$ state is shifted higher as $|V_+|$ decreases.
  }
\end{figure}

  Before discussing the subthreshold 1$^-_1$ state, let us explain how the subthreshold 2$^+_1$ state dominates the $S$-factor at $E_{c.m.}= 300$ keV.
  Figure~\ref{fig:E2} shows the schematic calculation of the $E$2 $S$-factors.
  The strength of the even-parity potential $V_+$ is varied.
  If the attraction of the nuclear potential $V_+$ becomes weaker, the sharp peak appears at low energies.
  And the peak energy is shifted higher as $V_+$ is getting weak.
  The shape of the resonance is not changed in the $S$-factors.
  This means that the resonant peak is the 2$^+_1$ bound state emerging from beneath the threshold.
  From figure~\ref{fig:E2}, we confirm that the $E$2 $S$-factors are enhanced by the the subthreshold 2$^+_1$ state.

\begin{figure}
  \begin{center}
    \includegraphics[width=0.6\linewidth]{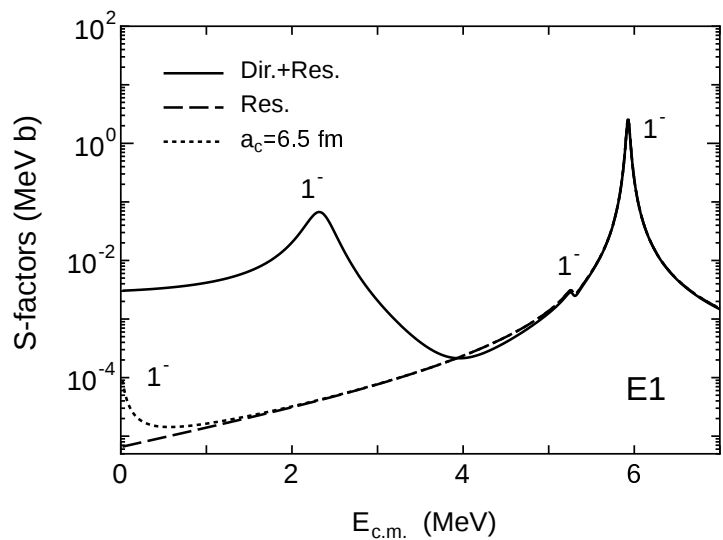}
  \end{center}
  \caption{\label{fig:E1}
    The $E$1 $S$-factors of $^{12}$C($\alpha$,$\gamma_0$)$^{16}$O.
    The solid curve is the result obtained from the potential model with the additional resonances.
    The dashed and dotted curves are the components of the additional resonances.
    The channel radius $a_c$ is presumed for the subthreshold state as the reference. (dotted curve)
  }
\end{figure}

  In contrast, the $E$1 $S$-factors are not enhanced by the subthreshold 1$^-_1$ state in the simple potential model \cite{Kat12,Kat08}.
  The solid curve in figure~\ref{fig:E1} is the $E$1 $S$-factors obtained from the potential model with the additional resonances.
  The dashed curves represent the component of the additional resonances.
  From this figure we find that the low-energy $S$-factors are not enhanced by the subthreshold 1$^-_1$ state.
  In the present model, the $\alpha$-particle width of the 1$^-_1$ state at low energies is reduced by $P(|E_R|)$, even if we surmise ${\it \Gamma}_{\alpha n}$ in equation~(\ref{eq:Gamma-a}). 
  We also find, from the comparison between two curves, that the direct-capture component dominates the $E$1 $S$-factors below $E_{c.m.}=3$ MeV.

  As an examination, we introduce the reduced width $\gamma^2_\alpha$ with the assumed channel radius $a_c$, in accordance with the $R$-matrix method.
  Instead of equation~(\ref{eq:Gamma-a}), we use the $\alpha$-particle width, defined by
  \begin{eqnarray}
    {\it \Gamma}_\alpha(E_{c.m.}) &=& 2P_L(E_{c.m.}) \gamma^2_\alpha,
    \label{eq:gamma-a-R}
  \end{eqnarray}
  for the subthreshold 1$^-_1$ state.
  $\gamma^2_\alpha$ is defined by the probability of $\alpha$-particle at the channel radius $a_c$.
  $P_L(E_{c.m.})$ is given by
  \begin{eqnarray}
    P_L(E_{c.m.}) &=& \left\{ \begin{array}{cl}
      \frac{\displaystyle{k_ia_c}}{\displaystyle{G_L^2(k_ia_c)+F_L^2(k_ia_c)}}  & (E_{c.m.} > 0) \\
                                            0    & (E_{c.m.} \le 0) \\
    \end{array},\right.
    \label{eq:pene}
  \end{eqnarray}
  where $F_L(k_ia_c)$ and $G_L(k_ia_c)$ are the regular and irregular Coulomb wave functions, respectively.
  $P_L(E_{c.m.})$ varies as equation~(\ref{eq:pene-low}) at low energies, $k_ia_c\approx0$.
  The dotted curve in figure~\ref{fig:E1} is calculated from equations~(\ref{eq:gamma-a-R}) and (\ref{eq:pene}) and ${\it \gamma}^2_{\alpha}=2.97$ keV \cite{Nacre2} for the 1$^-_1$ state.
  $a_c=6.5$ fm is used.
  If we use the width as in the $R$-matrix method, we find the tail of the subthreshold state at low energies.
  It however does not strongly interfere with the direct-capture component.
  We thus figure out that the resulting total $E$1 $S$-factor with equations~(\ref{eq:gamma-a-R}) and (\ref{eq:pene}) is almost identical to the result with equation~(\ref{eq:pene-low}).

  The $E$1 transition is hindered by the isospin selection rule \cite{Eis76}, and the absolute value of the $E$1 $S$-factors is reduced at low energies.
  Under the weak coupling, the additional narrow resonance cannot interfere with the 1$^-_2$ state ($E_{c.m.}=2.42$ MeV) belonging to the $\alpha$+$^{12}$C rotational band which has the relatively large amplitude.
  We also predict, for the same reason, that the $S$-factor of the 1$^-_3$ state at $E_{c.m.}=5.282$ MeV ($E_x=12.44$ MeV) is reduced if we use the given resonance width \cite{Til93}.

  In the $R$-matrix analyses, the {\it transparent} feature of the reactions is not taken into account.
  The $^{16}$O structure described by the relative motion between $\alpha$-particle and $^{12}$C might not be considered correctly.
  The large violation of the isospin selection rule is surmised to be allowed in the {\it internal} region.

  The resulting $S$-factors at $E_{c.m.}= 300$ keV are $S_{E1}\approx 3$ keV~b and $S_{E2}=152$ keV~b, respectively.
  They are approximately the same as our previous values \cite{Kat08}.

\subsection{$^{12}$C($\alpha$,$\gamma$)$^{16}$O reaction rates}

\begin{figure}
  \begin{center}
    \includegraphics[width=0.5\linewidth]{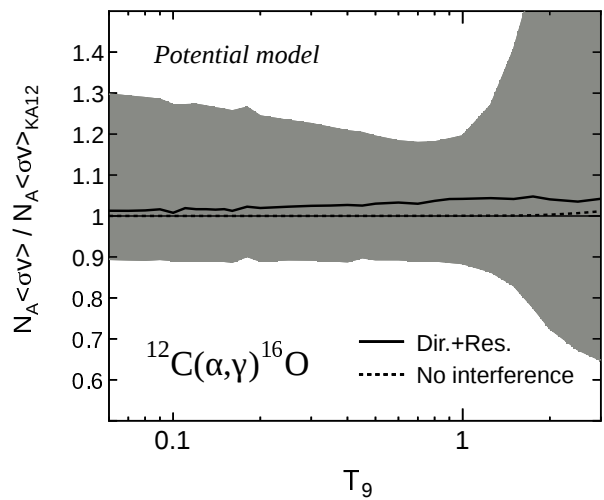}
  \end{center}
  \caption{\label{fig:rate}
    The thermonuclear reaction rates for the $^{12}$C($\alpha$,$\gamma$)$^{16}$O reaction.
    The reaction rates are displayed in the ratio to our recommended reaction rates (KA12) \cite{Kat12}.
    The solid curve is the result obtained from the numerical integration of equation~(\ref{eq:rate}).
    The dotted curve is the result with equation~(\ref{eq:rate-res}) for the resonances.
    The shade region is the uncertainties estimated from the variation of the model parameters.
    The difference from unity indicates the contribution of the additional resonances included in the present calculation.
  }
\end{figure}

  Figure \ref{fig:rate} compare the derived reaction rates with our recommended reaction rates (KA12) in \cite{Kat12}.
  The solid curve is the result displayed in the ratio to KA12.
  The difference from unity indicates the contribution of the additional resonances in the present calculation.
  The shade region is the uncertainties estimated from the variation of the model parameters \cite{Kat12}.
  We find, from the figure, that the additional resonances do not make the profound change in the reaction rates below $T_9=3$.
  They produce the difference below 5 \%.
  The dotted curve represents the reaction rates with equation~(\ref{eq:rate-res}).
  We find that the resonant component from equation~(\ref{eq:rate-res}) is very small.
  The dotted curve does not include the interference between the resonances and the direct-capture component.
  So, the increase in the reaction rate of the solid curve is caused by the weak interference between the components, rather than the resonant peak.
  The numerical integration of equation~(\ref{eq:rate}) below $T_9=3$ is in convergence at $E_{c.m.}=7$ MeV, satisfactorily.

  From figure~\ref{fig:rate}, we also find that the uncertainties of the reaction rates still remain at high temperatures, $T_9>1$.
  The primary reason of the uncertainties comes from the cascade transition through the excited state of $^{16}$O, as discussed in \cite{Kat12,Kat08}.
  They will be discussed in detail in the forthcoming paper.

\section{Summary}
\label{sec4}

  We have examined the additional resonant contribution to the potential model for the $^{12}$C($\alpha$,$\gamma$)$^{16}$O reaction.
  We have calculated the excitation function of $\alpha$+$^{12}$C elastic scattering below $E_{c.m.}= 8$ MeV, the low-energy astrophysical $S$-factors and the reaction rates of $^{12}$C($\alpha$,$\gamma$)$^{16}$O below $T_9=3$.
  In the present calculation, we use the parity-dependent real potential and include the known resonances as the Breit-Wigner form.

  The excitation functions and phase shifts of $\alpha$+$^{12}$C elastic scattering below $E_{c.m.}= 5$ MeV seem to be satisfactorily reproduced.
  The potential scattering appears to give the smooth trend of the excitation function and the single-particle potential resonances below $E_{c.m.}= 8$ MeV.
  The remaining rapid variation on the excitation function originates from the additional resonances.
  From the comparison, we find that the weak coupling to other reaction channels is good enough to describe in outline the structure of $^{16}$O below $E_x\approx 15$ MeV.

  The additional resonances complement the astrophysical $S$-factors obtained from the potential model.
  However, they do not give the large contribution to the derived $^{12}$C($\alpha$,$\gamma$)$^{16}$O reaction rates below $T_9=3$. (below 5\%)
  The $S$-factor at $E_{c.m.}=300$ keV is dominated by the $E$2 transition because of the subthreshold 2$^+_1$ state.
  The $E$1 is not enhanced by the subthreshold 1$^-_1$ state.
  The $S$-factors at $E_{c.m.}= 300$ keV are found to be $S_{E1}\approx 3$ keV~b and $S_{E2}=152$ keV~b, and they are approximately the same as the results of \cite{Kat08}.
  The simple potential model \cite{Kat08} describes the fundamental process of the radiative capture reactions.
  Compared with the direct-capture component, the additional $E$1 resonant component is small, and it does not strongly couple to the direct-capture process at low energies.

\ack
  The author thanks Professors K.~Langanke, G.~Mart\'inez-Pinedo, and I.J.~Thompson for their comments and hospitality during his visiting program.
  He also thanks Professors Y.~Ohnita and Y.~Sakuragi for their hospitality and encouragement and Professor Y.~Kond\=o for the early days of the collaboration.
  He is grateful to Professors M.~Arnould, A.~Jorissen, K.~Takahashi, and H.~Utsunomiya for their hospitality during his stay at Universit\'e Libre de Bruxelles (ULB).
  The part of this work has been supported by the Interuniversity Attraction Pole IAP 5/07 of the Belgian Federal Science Policy (Konan University -- ULB convention), and by the JSPS Institutional Program for Young Researcher Overseas Visits ``Promoting international young researchers in mathematics and mathematical sciences led by Osaka City University Advanced Mathematical Institute''.

\section*{References}

\clearpage
\Tables

\begin{table}
  \caption{\label{tb:bw}
    The $\alpha$+$^{12}$C resonant states of $^{16}$O \cite{Til93,Mar72,TI8} attached in the present calculation.
    The resonance energy $E_n$, the $\alpha$-particle width ${\it \Gamma}_{\alpha n}$, the $\gamma$-width to the ground state ${\it \Gamma}_{\gamma_0 n}$, and the total width ${\it \Gamma}_n$ are listed.
    $E_n$ is given in the center-of-mass energy.
    ${\it \Gamma}_{\alpha n}$ is indicated in the ratio to ${\it \Gamma}_n$.
  }
  \begin{indented}
  \item[]\begin{tabular}{cccccc}
    \br
    $n$ &  $E_n$ (MeV) & $J^\pi$ & ${\it \Gamma}_n$ (keV) & ${\it \Gamma}_{\alpha n}/{\it \Gamma}_n$ & ${\it \Gamma}_{\gamma_0 n}$ (eV) \\
    \mr
          1 & -0.045 &  1$^-$ &    |   &   |   & $7.9\times10^{-2}$ \\  
          2 &  2.683 &  2$^+$ &   0.625&  1.0  & $5.7\times10^{-3}$\\
          3 &  3.934 &  4$^+$ &   0.28 &  1.0  &   | \\
          4 &  4.357 &  2$^+$ &   73.  &  1.0  &  0.65\\  
          5 &  4.889 &  0$^+$ &    1.5 &  1.0  &   | \\
          6 &  5.282 &  1$^-$ &   99.  &  0.93 &  9.5\\   
          7 &  5.865 &  2$^+$ &  150.  &  1.0  &  0.7\\   
          8 &  5.928 &  1$^-$ &  130.  &  0.35 & 44.0\\   
          9 &  5.970 &  3$^-$ &  110.  &  0.82 &   | \\   
         10 &  6.098 &  3$^-$ &   26.  &  0.35 &   | \\
         11 &  6.720 &  4$^+$ &   75.  &  0.65 &   | \\
         12 &  6.873 &  0$^+$ &  200.  &  1.0  &   | \\
         13 &  6.975 &  3$^-$ &  750.  &  0.20 &   | \\
         14 &  7.461 &  4$^+$ &  487.  &  0.80 &   | \\
         15 &  7.646 &  6$^+$ &   70.  &  0.31 &   | \\
         16 &  7.908 &  0$^+$ &  166.  &  0.80 &   | \\
         17 &  8.249 &  3$^-$ &  133.  &  0.58 &   | \\
  \br
  \end{tabular}
  \end{indented}
\end{table}

\clearpage

\clearpage

\clearpage

\clearpage

\clearpage

\clearpage

\clearpage

\end{document}